\documentclass{elsart}

\usepackage{graphicx}
\usepackage{graphicx,amssymb}

\begin{document}

\begin{frontmatter}



\title{Neutron Background Studies for the CRESST Dark Matter Experiment}


\author{H. Wulandari\corauthref{cor1}},
\ead{Hesti.Wulandari@ph.tum.de} \corauth[cor1]{Corresponding
author. Tel: +49-89-28914416, Fax: +49-89-28912680.}
\author{J. Jochum},
\author{W. Rau},
\author{F. von Feilitzsch}

\address{Physikdept. E-15, Technische Universit\"{a}t M\"{u}nchen, James-Franckstr.- \\85747 Garching, Germany}

\begin{abstract}
The new detection concept applied for the direct WIMP search experiment CRESST II, which enables a
clear discrimination between electron recoils and nuclear recoils,
will leave neutrons as the main background.
This background will soon limit the sensitivity of the
experiment and therefore become an important issue for the next phase of
CRESST. We have performed a study based on Monte Carlo simulations
to investigate how neutrons from different origins affect CRESST
and which measures have to be taken to reach the
projected sensitivity.
\end{abstract}

\end{frontmatter}

\section{Introduction}
CRESST (Cryogenic Rare Event Search with Superconducting
Thermometers) is a direct WIMP (Weakly Interacting Massive
Particles) search experiment using low temperature detectors
\cite{angloher}. The experiment is performed at the Gran Sasso
laboratory in Italy, at a depth of 3600 m w.e. underground. In
such a rare event search, where one expects an event rate of the
order of less than 0.01 count/keV/kg/day, background must be
suppressed as much as possible. In spite of the employment of
passive background reduction techniques for CRESST, i.e. deep
underground site, efficient shielding against radioactivity of
surrounding rock, and the use of radiopure materials inside the
shielding, there is remaining background dominated by $\beta$ and
$\gamma$ emission from nearby radioactive contaminants that leads
to electron recoils in the detector. On the other hand WIMPs, and
also neutrons, produce nuclear recoils. Therefore, the sensitivity
of detection can be improved dramatically if, in addition to the
passive shielding, the detector itself is able to discriminate
electron recoils from nuclear recoils and actively reject them.
Such an active rejection technique is made possible in CRESST II
by simultaneous measurement of phonons and scintillation light
\cite{meunier}.

While a very efficient suppression of the electromagnetic
background can be achieved by the aforementioned methods, neutron-induced nuclear recoils can
not be discriminated from the WIMP-induced nuclear recoils. They
remain as a background and will limit the sensitivity of the experiment. It is obvious that CRESST
II and other current and future direct dark matter search experiments have
to cope with neutron background to reach high sensitivity.
Understanding this background and the way to overcome it, is
hence crucial for these experiments.

The neutron flux present in the CRESST setup at Gran Saso comes
from different origins:
\begin{enumerate}
\item Low energy neutrons induced by fission and
($\alpha$,n) reactions due to uranium and thorium activities in
the surrounding rock and concrete. They give the bulk to the total
flux in the laboratory.
\item Low energy neutrons induced by fission in the shielding material and the setup.
\item High energy neutrons induced by muons in the rock.
These neutrons could do spallation reactions in the experimental
shield and produce additional neutrons.
\item High energy neutrons induced by muons in the shielding material
(especially lead).
\end{enumerate}

Based on Monte Carlo simulations we have studied the contribution
of each neutron source to the expected background rate in CRESST
II. We will first discuss the background induced by low energy neutrons (section 2). Section 3 starts with the investigation of the muon flux and spectrum at the Gran Sasso laboratory. Following are the muon-induced neutron production underground (subsection 3.2), the resulting neutron spectrum in the laboratory and inside the experimental setup (subsection 3.3) and finally the recoil spectrum in the CRESST detectors expected from this background source. Section 4 compares the background contribution from the different sources and discusses further possible improvement.

\section{Background induced by low energy neutrons}
\subsection{Contribution of ($\alpha$,n) and fission neutrons from
the rock/concrete} In a previous paper \cite{wulandari} we have
reported a detailed study of the neutron flux at the Gran Sasso laboratory,
especially in hall A, where the CRESST setup is now located. From
this study we have obtained a more detailed neutron energy spectrum in hall A than was available in literature before. We have also shown that the neutron flux in the hall is actually dependent
on the humidity of the concrete layer in the laboratory: the flux
is higher when the concrete is dry than when it is wet. We have used the dry-concrete spectrum to determine the
background count rate expected in the CRESST CaWO$_{4}$ detectors,
because it would give a conservative estimate. Besides, we
also found that our calculation of the integral flux in hall A is
consistent with the measurement performed by Belli et al.
\cite{belli} if we assume that the concrete in the hall is dry.

To get a spectrum of neutron-induced recoils in the target we have
performed Monte Carlo simulations using MCNP4B
\cite{briesmeister}. No neutron moderator is installed in the
present setup of CRESST \cite{angloher} and a simplified geometry,
which consist of Pb/Cu shield, was used in the simulations. The
outer dimension of the lead shield is
130\,cm\,x\,130\,cm\,x\,136\,cm and the thickness is 20\,cm.
Inside the lead is a 15\,cm thick copper layer housing the experimental cavity, in which a single cube detector
crystal of 4\,cm\,x\,4\,cm\,x\,4\,cm is placed. Using this
geometry, our simulation with the dry-concrete spectrum gives 69
cts/kg/y in the energy range of 15-25 keV, while the spectrum from
\cite{belli} gives 47 cts/kg/y. The lower limit (15 keV) here is
determined by the threshold of the detector system of CRESST II.

In CRESST II, 60 GeV WIMPs with a cross section as claimed by the
DAMA collaboration in \cite{bernabei2} would give 55 cts/kg/y
between 15 and 25 keV. With the aforementioned contribution of
neutron background, it is therefore difficult for CRESST II to
check the DAMA evidence without a neutron
moderator shield.

If a neutron moderator of 50\,cm polyethylene is placed outside the lead shield,
6\,x\,10$^{-3}$ cts/kg/y are expected in the energy range of 15-25 keV
for the dry-concrete spectrum, while the simulation with the
spectrum from \cite{belli} gives 3.6\,x\,10$^{-2}$ cts/kg/y, which
is six times higher. This is due to the fact that in spite of the
agreement between our result and \cite{belli} for the total flux,
the detailed spectra are different. In addition, the results in \cite{belli}
are given as integral fluxes over quite large energy bins. In
sampling the energy in each bin we have assumed an equal
probability in the whole bin. This makes the spectrum harder than
it should be. On the other hand, we have given our results as
integral fluxes over smaller bins \cite{wulandari}, and therefore
more detailed. The higher number of neutrons with energies above 7
MeV in the spectrum from \cite{belli}, which can partly penetrate the moderator, give more counts in the detector. This fact indicates the importance of detailed knowledge of the neutron
spectrum at the experimental site.

\subsection{Neutrons produced by fission in the lead shield}
The use of radio pure material for the setup and
shield in rare event search experiments is mandatory. However, it is
still necessary to check, whether the low remaining activity is
really harmless.

We have performed a simulation to study the effect caused by
neutrons from the radio impurity in the lead shield of CRESST.
Only neutrons induced by spontaneous fission of $^{238}$U were
considered in our simulation, because the contribution of
($\alpha$,n) reactions in lead is not significant. We found that
neutrons from this origin would give 2 cts/kg/y in the energy
range of 15-25 keV for a $^{238}$U concentration of one ppb, which
is a factor of about 30 lower than the rate expected from low
energy neutrons from the rock for the simple setup without neutron
moderator. Hence, for the present CRESST setup a contamination of
a few ppb $^{238}$U is still acceptable. This situation changes
however, when a 50\,cm polyethylene shield is put in place so that
the contribution of low energy neutrons from the rock is
significantly reduced. Only a few ppt $^{238}$U is already a
limiting neutron source in this case. It is clear, that even a
very low contamination in the lead shield can still be dangerous
for the experiment, especially because it is very close to the
detector.

The typical amount of radio impurity in lead commonly used in rare event search experiments is obtained from the measurements performed by several groups. Allesandrello et al. \cite{allesandrello} reported a $^{238}$U
contamination of $<$ 2\,ppb in roman lead and $<$ 12\,ppb in low
activity lead. Assuming an equilibrium of
$^{238}$U with its daughter products, the EDELWEISS collaboration has found an
upper contamination limit of 0.7\,ppb $^{238}$U in the most recent measurements of its lead. Previous
measurements made on a different lead sample gave 0.1\,ppb \cite{gerbier}.

Additional neutron flux can be expected from the copper and
polyethylene shields. The contribution of the latter might be
negligible, because neutrons will be moderated by polyethylene.
To know the real contribution of neutrons from this origin, measurements of the contamination in the shielding materials used in CRESST are required.

\section{Muon-induced neutron background}
\subsection{Muon flux and energy spectrum}
To calculate differential and integral muon intensities at the
depth of the Gran Sasso laboratory a special code called SIAM
\cite{kudryavtsev} was used. In this code the differential muon
intensity underground was determined using the following
equation:
\begin{equation}
I_{\mu}(E_{\mu},X,\mbox{cos}\,\theta)=\int_{0}^{\infty}P(E_{\mu},X,E_{\mu0})
\frac{dI_{\mu0}(E_{\mu},\mbox{cos}\,\theta^{*})}{dE_{\mu0}}dE_{\mu0}
\label{eq:mu_intens}
\end{equation}
where
$\frac{dI_{\mu0}(E_{\mu},\mbox{cos}\,\theta^{*})}{dE_{\mu0}}$ is
the muon intensity at sea level at the zenith angle $\theta^{*}$:
\begin{equation}
\frac{dI_{\mu0}(E_{\mu},\mbox{cos}\,\theta^{*})}{dE_{\mu0}}\!=\!A
\frac{0.14E_{\mu}^{-\gamma}}{\mbox{cm}^{2}\mbox{s\,sr\,GeV}}\mbox{x}
\left\{\frac{1}{1\!+\!\frac{1.1E_{\mu0}\mbox{cos}\,\theta^{*}}{115\,\mbox{GeV}}}\!+\!
\frac{0.054}{1\!+\!\frac{1.1E_{\mu0}\mbox{cos}\,\theta^{*}}{850\,\mbox{GeV}}}\!+\!R_{c}\right\}
\label{eq:muspect_sealevel2}
\end{equation}
The relation between the zenith angle at the Earth's surface, $\theta^{*}$, and the zenith angle underground, $\theta$, was determined taking into account the curvature of the Earth. $R_{c}$ denotes the ratio
of prompt muons to pions. The parameters in
Eq.~(\ref{eq:muspect_sealevel2}) were taken either according to
Gaisser's parameterization \cite{gaisser} ($A\!=\!1,
\gamma\!=\,2.70$) which is modified for large zenith angles and
prompt muon flux \cite{aglieta}, or following the best fit to the
depth-vertical $\mu$ intensity relation measured by the LVD
experiment \cite{aglieta}. LVD reported the normalization constant
$A\!=\!1.84\pm 0.31$, $\gamma\!=\,2.77\pm 0.02$ and the upper
limit $R_{c}\leqslant 2$\,x\,10$^{-3}$ (95\% C.L.)
\cite{aglieta2}. $P(E_{\mu},X,E_{\mu0})$ is the probability for a
muon with energy $E_{\mu0}$ at the surface to have the energy
$E_{\mu}$ at depth $X$ \cite{kudryavtsev} and was obtained by
propagating muons with various energies at the Earth's surface
using MUSIC (Muon Simulation Code) \cite{antonioli2}.

In this work we have taken $10^{-4}$ for the ratio of prompt muons
to pions, well below the upper limit given by LVD \cite{aglieta2}.
To calculate the integral muon intensity, an integration of
Eq.~(\ref{eq:mu_intens}) over $dE_{\mu}$ was carried out. A
further integration over cos$\,\theta$ gives the global intensity
for a spherical detector.

The absolute muon intensity underground depends in fact on the
surface relief. Gran Sasso has a very complex mountain profile,
that makes it difficult to predict the muon flux without precise
information on the slant depth distribution. In this work a flat
surface was assumed as approximation. In the near future, a
further study is planned, taking the detailed mountain profile of
Gran Sasso into account. There might not be a big difference in
the neutron production -- the simulation results will after all be
normalized to the muon flux measured in Gran Sasso
($1\mu/$h/m$^{2}$) -- but it may be important for
simulations of the muon veto.

\subsection{Muon-induced neutron production at Gran Sasso}
We have used FLUKA \cite{fasso} to simulate neutron production by
muons. The two muon energy spectra underground discussed in the
previous subsection have been used in the simulations, i.e. the
one obtained with the parameters according to Gaisser's
parameterization and the other following the LVD best fit. In
addition, simulations have also been performed for mono energetic
muons of 270 GeV, which is the mean muon energy at the depth of
Gran Sasso \cite{aglietta3}.

The neutron production has been simulated for materials relevant
for CRESST, i.e. Gran Sasso rock and concrete, lead, copper, and
polyethylene. The compositions of Gran Sasso rock and concrete are taken from \cite{catalano} and \cite{wulandari} respectively. Neutron
production rates obtained from the simulations
with the spectrum following the LVD best fit are shown in Table~\protect\ref{nprod_LVD} for three different processes: muon spallation, hadronic shower and electromagnetic shower. The rates are in agreement
with the results of simulations with the spectrum using Gaisser's
parameterization and also with the results from mono energetic 270
GeV muons. These data show that the neutron production rate at Gran
Sasso is dominated by secondary processes, i.e. hadronic and
electromagnetic showers. It can also be seen that neutron
production increases with the average atomic weight $\left\langle A \right\rangle$ of the target. The
production rates in rock and concrete are about the same due to
their similar average atomic weights.

There are some measurements on the neutron production rate by
muons in lead that can be used as a comparison for the result of
this work. Gorshkov measured neutron production by
muons at several depths underground for several elements including lead (see \cite{gorshkov}
and reference therein).
At 800 m w.e. he found an average neutron production per nucleon in lead $\overline{m\sigma}/A=300\,\mbox{x}\,10^{-29}\,\mbox{cm}^{2}/\mbox{Pb
nucleus}$, where $m$, $\sigma$ and $A$ denote multiplicity, cross section and atomic weight respectively. This number
multiplied by the Avogadro's number gives the neutron
production of 1.81\,x\,10$^{-3}$ neutrons/$\mu$/(g/cm$^{2}$).
The mean energy at this depth is approximately 110\,GeV. Using the law that the neutron production rate goes like $E_{\mu}^{0.75}$ as measured for liquid
scintillator \cite{zatsepin,khalchukov,wang} would give a neutron
production rate at 270\,GeV (Gran Sasso mean muon energy) of
3.55\,x\,10$^{-3}$ neutrons/$\mu$/(g/cm$^{2}$), which is in
reasonable agreement with the simulation in this work.

Bergamosco et al. performed an experiment on neutron production by
muons in lead at Mont Blanc (4300 m w.e, or 5200 m w.e according
to the LSD experiment \cite{aglietta4}). They reported a product of
multiplicity and cross section,
$\overline{m}\sigma=(400\pm150)\,\mbox{x}\,10^{-26}\,\mbox{cm}^{2}/\mbox{Pb
nucleus}$, that leads to a neutron production of $1.16\,$x\,10$^{-2}$ neutrons/$\mu$/(g/cm$^{2}$) at
this depth. If the mean muon energy at Mont Blanc is assumed to be
385\,GeV as reported by LSD \cite{aglietta4}, the $E_{\mu}^{0.75}$
law applied to Gorshkov's result would give a neutron production
rate in lead at Mont Blanc as $4.63\,$x\,10$^{-3}$
neutrons/(g/cm$^{2}$), which is almost three times lower than
Bergamosco's result. The $\overline{m}\sigma$ reported by
Bergamosco is, in fact, about three times higher than the
theoretical prediction and extrapolation from several other
experiments \cite{bergamosco}. That means, if his measurement was
off by a factor of three, the neutron production would be in
reasonable agreement with our and Gorshkov's results.

Recent measurements at CERN to determine the neutron production
rate in lead, copper and carbon by a 190\,GeV muon beam
(experiment NA55) were performed with thin targets and the results
are reported at certain scattering angles only \cite{chazal}.
Therefore this information can unfortunately not easily be used to
check our results.

\subsection{Energy spectrum of neutrons entering the experimental hall
and the Detector Area} To determine the energy spectrum of
neutrons from the rock entering the laboratory hall, the muon
spectrum with parameters following the LVD best fit was used. It
is clear that the number of neutrons entering the hall depends on the thickness of the rock used in the simulations. A
very large thickness needs too much computing time, whereas a too
small thickness will underestimate the particle yield. It was
found in our simulation of neutron production in Gran Sasso rock,
that cascades were well developed and that the equilibrium between
neutron and muon flux was reached after muons had crossed about
6-7 meter of Gran Sasso rock.

Additional information comes from the results of Monte Carlo
simulations performed by Dementyev \cite{dementyev}. He found that
the typical depth of Gran Sasso rock for hadrons with energies
above 200\,MeV is 90\,cm, which means that 96\% of the neutron flux
entering the hall is produced at a depth of up to about 3 meter
behind the rock surface.

In this work, muons were generated at the surfaces of a cube of
rock with a size of 20\,m\,x\,20\,m\,x\,20\,m. Inside the rock
cube, the experimental hall was taken to be of a size of
6\,x\,6\,x\,5\,m$^{3}$. The top of the hall was placed 10\,m below
the top of the rock cube. This should be the optimal depth to allow
the cascades to develop and to let neutrons produced in the
last 3\,-\,4 meters of rock overburden enter the hall. The
size of the experimental hall used in these simulations was
chosen smaller than the real hall at the Gran Sasso laboratory to save computing time.
But some test simulations have been
done to ensure that the results do not change significantly if a
larger size is used. To sample muon energy and angular
distribution the code MUSUN (Muon Simulation Underground)
\cite{kudryavtsev} was used.

\begin{figure}[p]
\begin{center}
\includegraphics*[width=3.4in]{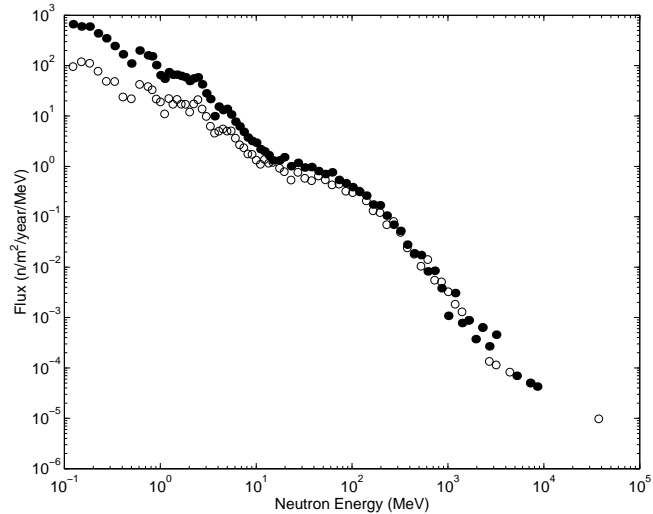}
\end{center}
\caption{Flux of muon-induced neutrons entering the Gran Sasso
hall obtained from simulations in this work,
without $(\mbox{\large{$\circ$}})$ and
with $(\mbox{\large{$\bullet$}})$ back scattering.
\label{flux_mu_comp}}
\end{figure}

\begin{figure}[p]
\begin{center}
\includegraphics*[width=3.4in]{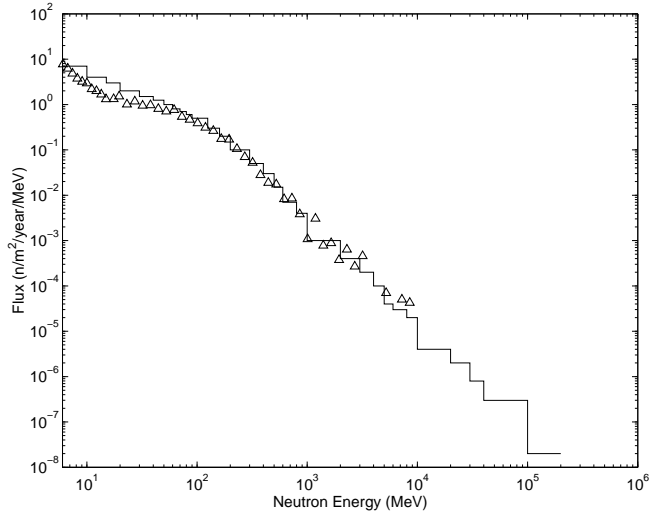}
\end{center}
\caption{Comparison of the spectra of muon-induced neutrons entering the
Gran Sasso hall: Dementyev \protect\cite{dementyev} (solid line)
and this work (with back scattering, triangles).
\label{dement_vs_mine}}
\end{figure}

Two different cases have been considered: all neutrons were absorbed immediately after entering the hall (that is without back scattering) and neutrons were scattered by the rock surrounding the hall (with back scattering).
Figure~\protect\ref{flux_mu_comp} shows the neutron energy spectra
at the boundary of rock and hall for these two cases. The total flux of neutrons above 1 MeV entering the hall
is 4.27\,x\,10$^{-10}\,$n/cm$^{2}$/s
(135\,n/m$^{2}$/year) for the first and 8.53\,x\,10$^{-10}\,$
n/cm$^{2}$/s (269\,n/m$^{2}$/year) for the second case. In
Figure~\protect\ref{dement_vs_mine} our spectrum with back scattering is shown together with the spectrum reported by
Dementyev \cite{dementyev}. The two spectra are in
agreement at high energy, but for the energy range between
6\,-\,60\,MeV Dementyev's flux is higher.

\begin{figure}[p]
\begin{center}
\includegraphics*[width=5.2in]{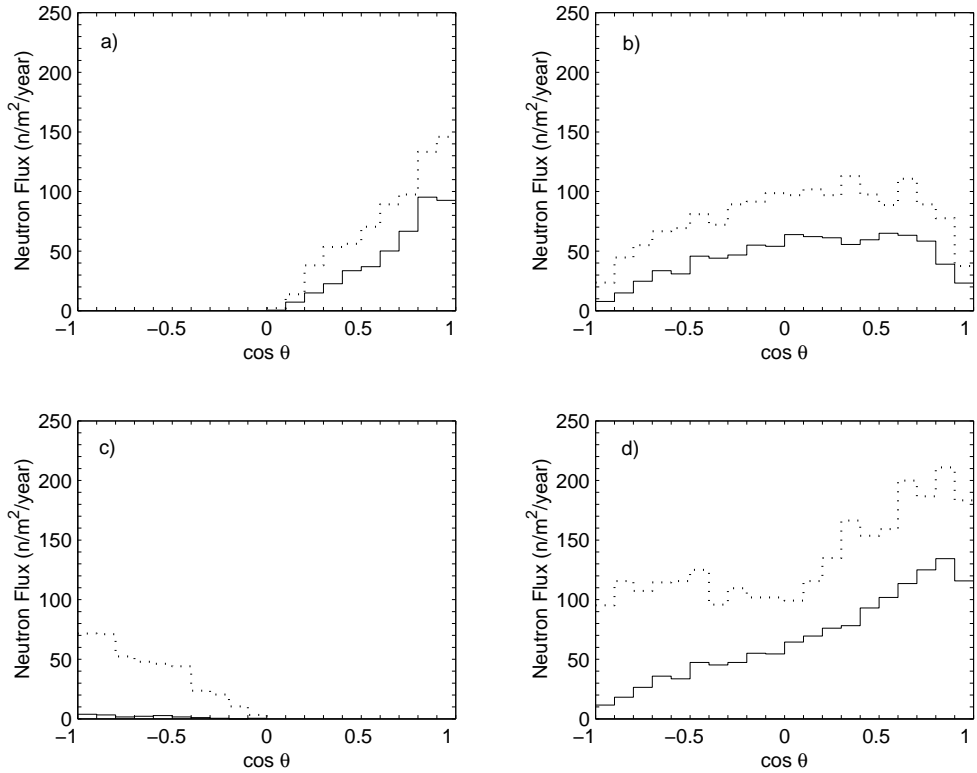}
\end{center}
\caption{Angular distribution of neutrons  with
energies above 1 MeV entering the hall from
a) the roof, b) the walls, c) the floor and d) anywhere for the case without back scattering (solid
lines) and with back scattering (dotted lines).
 \label{angular_dist}}
\end{figure}

\begin{figure}[p]
\begin{center}
\includegraphics*[width=5.2in]{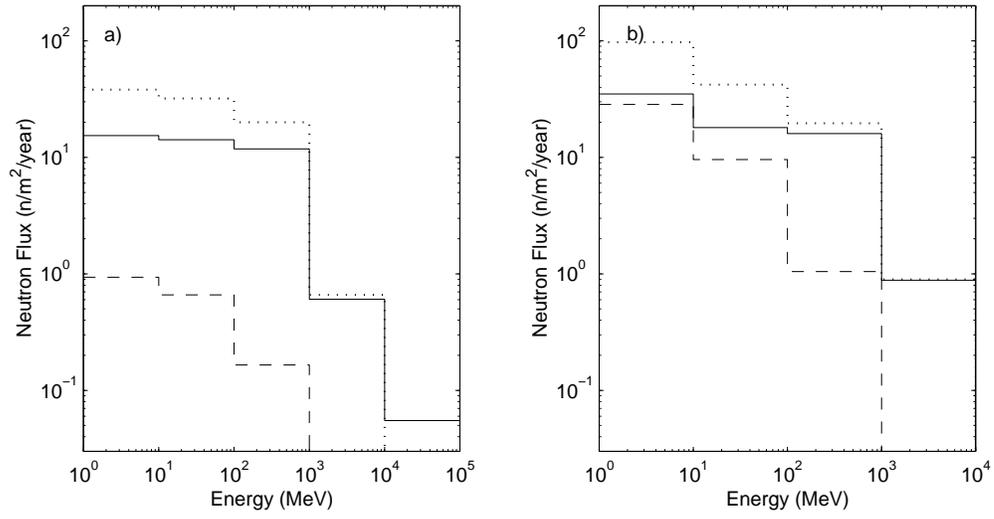}
\end{center}
  \caption{Energy distribution of
neutrons entering the hall from the roof (solid lines), the walls
(dotted lines) and the floor (dashed lines) for the cases a)
without back scattering and b) with back scattering.
 \label{n_energy_hall}}
\end{figure}

The angular distributions of neutrons entering the hall with
energies above 1\,MeV are shown  in Figure~\protect\ref{angular_dist} separately for the roof, the
floor, the walls and total for the cases with and without back
scattering. Almost all neutrons
entering the hall from the floor are back scattered neutrons.
Energy spectra of neutrons with energies above 1 MeV entering the
hall from the roof, the wall and the floor are shown in
Figure~\protect\ref{n_energy_hall} for the cases without and with
back scattering. High energy neutrons come mainly from the roof.
The number of lower energy neutrons increases due to scattering.

\begin{figure}[p]
\begin{center}
\includegraphics*[width=4.in]{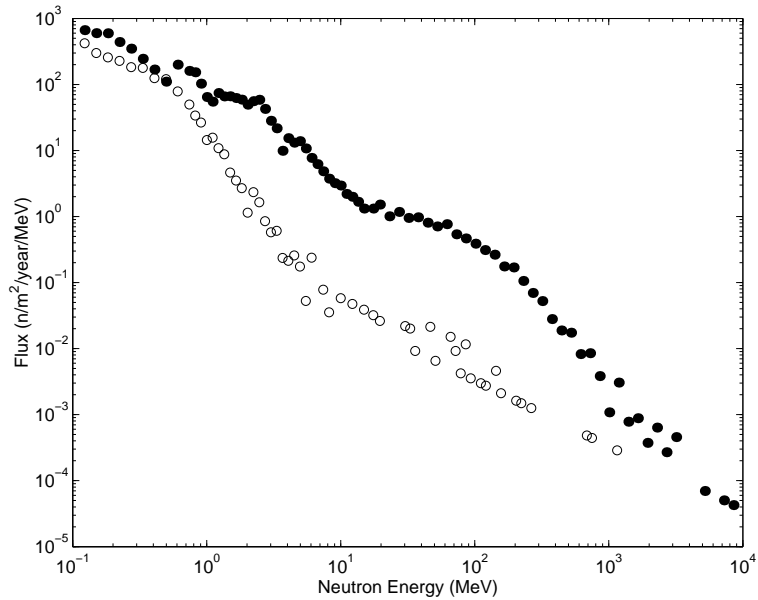}
\end{center}
\caption{The flux at the boundary between the shield and the
detector area inside the shield ($\mbox{\large{$\circ$}}$). As a
comparison the flux of neutrons entering the hall (including back
scattering) is shown ($\mbox{\large{$\bullet$}}$).}
\label{shield_effect}
\end{figure}

In Figure~\protect\ref{shield_effect} the neutron flux at the boundary
between the shield and the detector area inside the shield is
shown. The neutrons here are produced by muons both in the rock
and in the shielding materials used in CRESST. The flux below 1\,MeV
comes mainly from neutrons produced by muons in the lead shield.

\subsection{Recoil spectrum and count rate of muon-induced neutron background}
While the background induced by muons in the shield can efficiently be removed by a muon veto system,
the neutrons produced by muons in the rock can not be suppressed easily. Therefore, we investigate
these two contributions separately.

Because FLUKA does not treat individual nuclear recoils, MCNPX \cite{waters} and MCNP4B \cite{briesmeister}
were employed to study the contribution of muon-induced neutrons in the rock to the expected background rate
in the CRESST CaWO$_{4}$ detector. The simple experimental setup as described in section 2.1 with 50\,cm polyethylene shield was placed
inside the hall and neutrons (the spectrum with back scattering) were generated at the hall's surface and
transported further down
to the detector level to eventually get the recoil spectrum.
In the experimental setup these high energy neutrons produced
additional neutrons. The surrounding rock was replaced with
vacuum, to prevent neutrons from being scattered again by the rock.
The count rate in the energy range of (15$\,-\,$25)\,keV is
8.81\,x\,10$^{-4}\,$cts/kg/day. This result is higher than the
rates for low energy neutrons from activity of the surrounding
rock and concrete with the same polyethylene thickness.

To determine the contribution of muon-induced neutrons in the
experimental setup, muons following the LVD best fit were
generated at the hall's surface and the simple setup with
50\,cm polyethylene shield was placed inside the hall. Muons were
transported inside the hall and through the experimental setup with FLUKA.
The rock surrounding the experimental hall was replaced by vacuum,
to avoid muon-induced neutron production therein.
Neutrons produced  by muons in the experimental setup and entering
the experimental cavity (passing through the boundary between the
inner surface of the copper shield and the experimental cavity)
were then transported further with MCNPX and MCNP4B to get the recoil
spectrum in the detector crystal. The count rate in
(15$\,-\,$25)\,keV range is 2.78\,x\,10$^{-3}\,$cts/kg/day. This
rate is higher than that from low energy neutrons from the
rock/concrete after being moderated by 50\,cm polyethylene and
even higher than that rate from high energy neutrons from the rock.

\section{Comparison of neutron background from different sources}
\begin{figure}[p]
\begin{center}
\includegraphics*[width=3.3in]{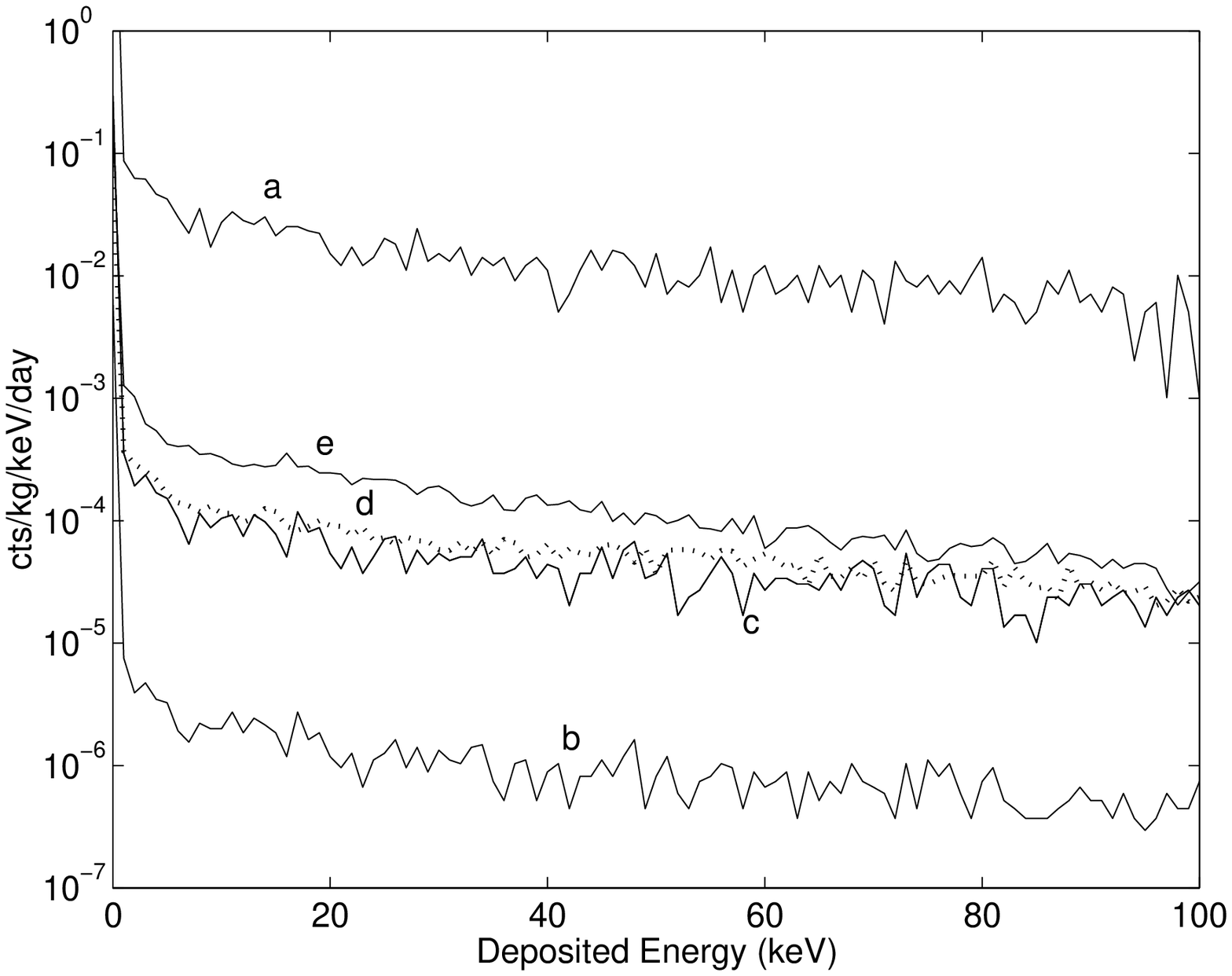}
\end{center}
\caption{Recoil spectra in a CaWO$_{4}$ detector induced by
neutrons from different origins: (a) low energy neutrons from the
rock/concrete, no neutron moderator, (b) low energy neutrons from
the rock/concrete after being moderated by 50\,cm polyethylene,
(c) low energy neutrons from fission reactions of 0.1 ppb
$^{238}$U in the lead shield, (d) high energy neutrons induced by
muons in the rock and (e) high energy neutrons induced by muons in
the experimental setup. \label{recoil_all}}
\end{figure}

In Figure~\protect\ref{recoil_all} the recoil spectra in the
CaWO$_{4}$ detector induced by neutrons from different origins are
shown. The recoil spectra induced by low energy neutrons from the
rock/concrete have been obtained with the dry-concrete spectrum. The spectrum for neutrons from fission reactions in the lead shield here is obtained by taking a $^{238}$U contamination of 0.1\,ppb.

The remaining neutron flux with the neutron moderator installed is
dominated by neutrons induced by muons in the lead shield. This
background would limit the sensitivity of the experiment for the
WIMP-nucleon cross section to about $10^{-7}\,$pb. As
aforementioned, it is possible to suppress this neutron background
by a muon veto system. For CRESST II such a system is planned in
addition to a neutron moderator. This will enable CRESST II to
reach the projected sensitivity for the WIMP-nucleon cross section
of below $10^{-7}\,$pb. The muon veto will be placed inside the
polyethylene shield and will have an efficiency of more than 90\%.
However, the muon veto will reduce the neutron background only by
a factor of three, unless high energy neutrons from the rock can
be overcome. Neutrons scattered in more than one detector can be
rejected as background, because WIMPs do not scatter multiply.
Therefore the neutron background will be further reduced and
multiple scattering can also be engaged to determine the remaining
single scatter neutron background. This technique has already been
successfully applied for the CDMS experiment \cite{abrams}.
Simulations with an array of detectors will be done in the near
future to investigate, which sensitivity level can be reached by
this technique.

\section{Conclusions}
We have discussed the contribution of different neutron background
sources relevant for CRESST. The flux of low energy neutrons from
the surrounding rock/concrete can be reduced efficiently by a
hydrogen-rich material like polyethylene. For the CRESST II setup
a polyethylene shield (35-50 cm thick) is advisable. This will
reduce the background count rate in the CaWO$_{4}$ detector by
more than three orders of magnitude. Then the background will be
dominated by neutrons from other origins. To reach the projected
sensitivity, a muon veto is planned for  CRESST. Multiple
scattering should be studied and the radio impurity of the
shielding materials need to be measured to determine a more
realistic contributions of muon-induced neutrons in the rock and
fission-induced neutrons in the shielding materials.

\section{Acknowledgements}
We are thankful to all members of the working group ETNo$\mu$SiQ
(European Team for Neutron and $\mu$ Shield Qualification), especially to Dr. Vitaly Kudryavtsev
at the University of Sheffield for allowing us to use the codes SIAM, MUSUN and MUSIC and for any
assistance in the muon simulations, and to Dr. Gilles Gerbier at DAPNIA, CEA/Saclay for valuable
discussions on low energy neutrons. H. Wulandari thanks the \textit{Deutscher Akademischer Austausch Dienst}
(DAAD) for the financial support of her PhD work.

\begin{table}[p]      
\caption{Neutron production rate from muons with the LVD best fit
spectrum in several materials.} {\tabcolsep2pt
\begin{tabular}{|c|c|c|c|c|c|}
\hline Material & Average atomic weight & \multicolumn{4}{|c|}{Neutron
production rate (10$^{-5}$
n/$\mu$/(g/cm$^{2}$))}\\
\cline{3-6}
 & $\left\langle A \right\rangle$ &$\mu$-spallation & Hadronic shower & E.m. shower & Total\\
\hline\hline
Lead &207.20 & 10.13 &239.45 &178.68 &428.26\\
Copper &63.55 &4.22& 74.95 &41.04 &120.20\\
LNGS rock &22.87 &1.79 &26.42 &7.37 & 35.59\\
LNGS concrete &20.50 &1.86 &25.40 &5.70 &32.96 \\
Polyethylene &10.40 &1.23 &16.32 &6.27 &23.82\\
\hline
\end{tabular}
\label{nprod_LVD}}
\end{table}

\begin{thebibliography}{99}
\bibitem{angloher}
G. Angloher et al., CRESST Collaboration, Astro. Part. Phys. 18
Vol. 1 (2002) 43.

\bibitem{meunier}
P. Meunier et al., Appl. Phys. Lett. 75 (1999) 1335.

\bibitem{wulandari}
H. Wulandari et al., hep-ex/0312050.

\bibitem{belli}
P. Belli et al., Il Nuovo Cim. 101 A, N.6 (1989) 959.

\bibitem{briesmeister}
J.F. Briesmeister, Ed., ``MCNP-A General Monte Carlo N-Particle
Code, Version 4B", LA-12625-M, Los Alamos National Laboratory
(March 1997).

\bibitem{bernabei2}
R. Bernabei et al., ROM2F/2000/01.

\bibitem{allesandrello}
A. Allesandrello et al., Nucl. Inst. Meth. in Phys. Res.
B61 (1991)106.

\bibitem{gerbier}
G. Gerbier, private communication.

\bibitem{kudryavtsev}
V.A. Kudryavtsev, private communication.

\bibitem{gaisser}
T.K. Gaisser, ``Cosmic Rays and Particle Physics", Cambridge
University Press, (1990)

\bibitem{aglieta}
M. Aglietta et al., Phys. Rev. D 58 (1998) 092005.

\bibitem{aglieta2}
M. Aglieta et al. (The LVD-Collaboration), Phys. Rev. D 60 (1999)
112001.

\bibitem{antonioli2}
P. Antonioli, V.A. Kudryavtsev, E.V. Korolkova and N.J.C. Spooner, Phys. Lett. B 471 (1999) 251.

\bibitem{fasso}
A. Fasso, A. Ferrari and P.R. Sala, in Proceedings of the Monte
Carlo 2000 Conference, (Lisbon, October 23-26, 2000), Eds. A. King,
F. Barao, M. Nakagawa, L. Tavora, P. Vaz, Springer-Verlag, Berlin
(2001) 159; A. Fasso, A. Ferrari, J. Ranft and P.R. Sala, ibid.
995.

\bibitem{aglietta3}
M. Aglietta et al., in Proc. 26th Intern. Cosmic Rays
Conf., V.2 (1999) 44; hep-ex/9905047.

\bibitem{catalano}
P.G. Catalano et al., Mem. Soc. Geol. It., 35 (1986)
647.

\bibitem{gorshkov}
G.V. Gorshkov et al., Sov. J. Nucl. Phys., Vol.18, No.1 (1974) 57.

\bibitem{zatsepin}
G.T. Zatsepin and O.G. Ryazhkaya, in Proceedings of the IX
International Conf. on Cosmic Rays, Vol.3 (London 1966) 987.

\bibitem{khalchukov}
F.F. Khalchukov et al., Il Nuovo Cim. 18C, N.5 (1995)
517.

\bibitem{wang}
Y.-F. Wang et al., Phys. Rev. D 64 (2001) 013012.

\bibitem{aglietta4}
M. Aglietta et al., Nuovo Cim. Soc. It. Fis. C 12 (1989)
467.

\bibitem{bergamosco}
L. Bergamosco, S. Costa and P. Picchi, Il Nuovo Cim. 13A, N.2
(1973) 403.

\bibitem{chazal}
V. Chazal et al., Nucl. Inst. Meth. in Phys. Res. A 490
(2002) 334.

\bibitem{dementyev}
A. Dementyev et al., Nucl. Phys. B (Proc. Suppl.) 70
(1999) 486. Nucl. Inst. Meth. in Phys. Res. A 314 (1992) 380.

\bibitem{waters}
L.S. Waters, ed. 1999, MCNPX User's Manual -
Version 2.1.5, Los Alamos National Laboratory Report,
TPO-E83-G-UG-X-00001.

\bibitem{abrams}
D. Abrams et al., Phys. Rev. D 66 (2002) 122003.

\end{thebibliography}
\end{document}